\documentclass[letterpaper]{article}
\usepackage{spconf,amsmath,graphicx}

\usepackage[noadjust]{cite}

\usepackage{amsmath,amssymb,bm}
\usepackage{url}
\usepackage{hyperref}
\usepackage[dvipsnames]{xcolor}
\newcommand\myshade{70}
\colorlet{mywholecolor}{MidnightBlue}
\hypersetup{
  linkcolor  = mywholecolor!\myshade!black,
  citecolor  = mywholecolor!\myshade!black,
  urlcolor   = mywholecolor!\myshade!black,
  colorlinks = true,
}

\renewcommand{\bm}[1]{\textbf{#1}}

\newcommand{\modelname}{MOVE}
\newcommand{\tabspace}{\vspace{-5mm}}

\newcommand{\footnotesizeq}{\fontsize{7.5pt}{9pt}\selectfont}

\usepackage{tikz}
\usepackage{textcomp}
\usepackage{hyperref}
\usepackage{lipsum}
\newcommand\copyrighttext{%
  \footnotesizeq \textcopyright Copyright 2020 IEEE. Published in the IEEE 2020 International Conference on Acoustics, Speech, and Signal Processing (ICASSP 2020), scheduled for 4-9 May, 2020, in Barcelona, Spain. Personal use of this material is permitted. However, permission to reprint/republish this material for advertising or promotional purposes or for creating new collective works for resale or redistribution to servers or lists, or to reuse any copyrighted component of this work in other works, must be obtained from the IEEE. Contact: Manager, Copyrights and Permissions / IEEE Service Center / 445 Hoes Lane / P.O. Box 1331 / Piscataway, NJ 08855-1331, USA. Telephone: + Intl. 908-562-3966.}
\newcommand\copyrightnotice{%
\begin{tikzpicture}[remember picture,overlay]
\node[anchor=south,yshift=10pt] at (current page.south)
{\fbox{\parbox{\dimexpr\textwidth-\fboxsep-\fboxrule\relax}{\copyrighttext}}};
\end{tikzpicture}%
}

\title{ACCURATE AND SCALABLE VERSION IDENTIFICATION USING MUSICALLY-MOTIVATED EMBEDDINGS}
\name{Furkan Yesiler$^{\dagger}$ \qquad Joan Serr{\`a}\sthanks{Work done while at Telef\'onica Research, Barcelona.}$^{\ddagger}$ \qquad Emilia G{\'o}mez$^{\dagger\mathsection}$}
			\address{$^{\dagger}$ Music Technology Group, Universitat Pompeu Fabra, Barcelona, Spain \\
			    $^{\ddagger}$ Dolby Laboratories, Barcelona, Spain \\
			    $^{\mathsection}$ Joint Research Centre, European Commission, Sevilla, Spain} 

\begin{document}
\ninept
\maketitle
\copyrightnotice
\vspace{-5mm}
\begin{abstract}
%The version identification (VI) task consists of automatically detecting recordings that correspond to the same musical work. Current VI systems  In this paper, we propose MOVE, a musically-motivated method for accurate and scalable version identification. MOVE achieves state-of-the-art performance on two publicly available benchmark sets by improving the previous work with a useful input representation, a novel technique for temporal content summarization, a non-parametric batch normalization for obtaining a standardized latent space, and a VI-specific data augmentation technique. The model is trained with a triplet loss, using an online hard triplet mining strategy. Furthermore, we perform an ablation study to point out the significance of our design decisions, and we also study the relation between the embedding dimensionality and the model performance.
The version identification (VI) task deals with the automatic detection of recordings that correspond to the same underlying musical piece. Despite many efforts, VI is still an open problem, with much room for improvement, specially with regard to combining accuracy and scalability. In this paper, we present \modelname, a musically-motivated method for accurate and scalable version identification. \modelname\ achieves state-of-the-art performance on two publicly-available benchmark sets by learning scalable embeddings in an Euclidean distance space, using a triplet loss and a hard triplet mining strategy. It improves over previous work by employing an alternative input representation, and introducing a novel technique for temporal content summarization, a standardized latent space, and a data augmentation strategy specifically designed for VI. In addition to the main results, we perform an ablation study to highlight the importance of our design choices, and study the relation between embedding dimensionality and model performance.
\end{abstract}
\section{Introduction}
\label{sec:intro}

Version identification (VI) commonly refers to the task of determining, by computational means, whether two audio renditions correspond to the same underlying musical composition~\cite{serra2011}. Being more challenging than traditional audio fingerprinting~\cite{Cano05VLSI}, VI goes beyond near-exact duplicate detection to embrace additional perceptual differences that, despite having a contrasting imprint in the signal, convey the same musical entity~\cite{grosche2012}. Such is the case of changes in instrumentation, musical key, tempo, timing, structure, or lyrics, to name a few~\cite{serra2011}. Besides digital rights management, VI has application to music organization, retrieval, navigation, and understanding.

Traditional VI systems generally approach the task with a pipeline consisting of three main stages~\cite{osmalskyj2017}. Firstly, as many other content-based retrieval methods, VI systems use feature extraction to obtain relevant information from the audio signal. Representations like predominant melody, pitch class profiles (PCP), or the constant-Q transform (CQT) have proven useful for this initial step~\cite{marolt2006, kurth2008, humphrey2013}. Secondly, traditional VI systems use various post-processing strategies for achieving transposition, tempo, timing, or structure invariance~\cite{ellis2007, tsai2016, serra2009}. Thirdly, for estimating similarity between pairs of songs, VI systems use segmentation strategies or local alignment methods, which also introduce invariance with regard to musical piece structure~\cite{serra2009, gomez2006, tralie2017cover}. Further approaches have explored combining the information obtained from different features and/or different alignment schemes with early or late fusion techniques~\cite{foucard2010, tralie2017cover, chen2018}. These, together with some previous solutions, achieve good performance in different evaluation contexts but, nonetheless, have difficulties in scaling to datasets above tens of thousands of songs~\cite{yesiler2019}. With the release of the SHS dataset~\cite{millionsong}, researchers explored scalable approaches based on audio hashprints, the 2D Fourier transform, or motif-finding strategies~\cite{tsai2016, bertin2012, silva2016}, but those achieved a limited success compared to their predecessors.

Recent deep learning approaches for VI aim to provide systems that are both accurate and scalable. In general, they focus on learning accurate, low-dimensional embeddings of recordings for, later, estimating similarities with basic distance metrics, with the intention to exploit existing scalable nearest-neighbor libraries. Xu~et~al.~\cite{xu2018} and Yu~et~al.~\cite{yu2019} train their convolutional networks in a multi-class classification fashion, where each version group (or clique) is considered a unique class, and use PCP and CQT as their inputs, respectively. For evaluation, they use the representations obtained from the penultimate layer of their network as embeddings. Beyond classification-based strategies, deep metric learning approaches with contrastive and triplet losses are becoming popular for VI. Qi~et~al.~\cite{qi2018} use a convolutional network with PCPs as input and a triplet loss as the objective function. As an alternative to using PCP variants, Doras~\&~Peeters~\cite{doras2019} use a 2D predominant melody representation as input to their convolutional network, which is also trained with a triplet loss but using an online semi-hard triplet mining strategy.

%In this paper, we focus on obtaining an accurate and computationally efficient method that improves the state of the art for the CSI task, and our contributions are X-fold. 

In this paper, we propose a music embedding method that allows for both accurate and scalable VI. We call it \modelname: musically-motivated version embeddings. \modelname\ achieves state-of-the-art results on two publicly-available benchmark sets and, since it is based on Euclidean distances, allows for efficient retrieval and indexing using existing libraries. The architecture of \modelname\ introduces a number of improvements, including (1)~a relatively novel input representation that has not been explored in the context of deep metric learning for VI, (2)~a multi-channel adaptive attention mechanism that is an alternative to previously-used temporal aggregation strategies, and (3)~a non-parametric batch normalization at the last layer to yield a standardized embedding space. The training of \modelname, like other recent VI systems, is done with a triplet loss. However, in contrast to those, it uses an online hard triplet mining strategy. In order to learn invariances with respect to the modifiable musical characteristics, \modelname\ is trained with a VI-specific data augmentation strategy. 
%In order to justify our design and to provide better insight regarding \modelname's performance, we perform an ablation study and also investigate the relation between such performance and embedding dimensionality. 
To gain insight, we perform an ablation study and also investigate the role of embedding dimensionality. 
To enable further research, we evaluate our method on publicly-available datasets and provide our code at \url{https://github.com/furkanyesiler/move}.

\begin{figure*}[tb!]
  \centering
  \includegraphics[width=\textwidth]{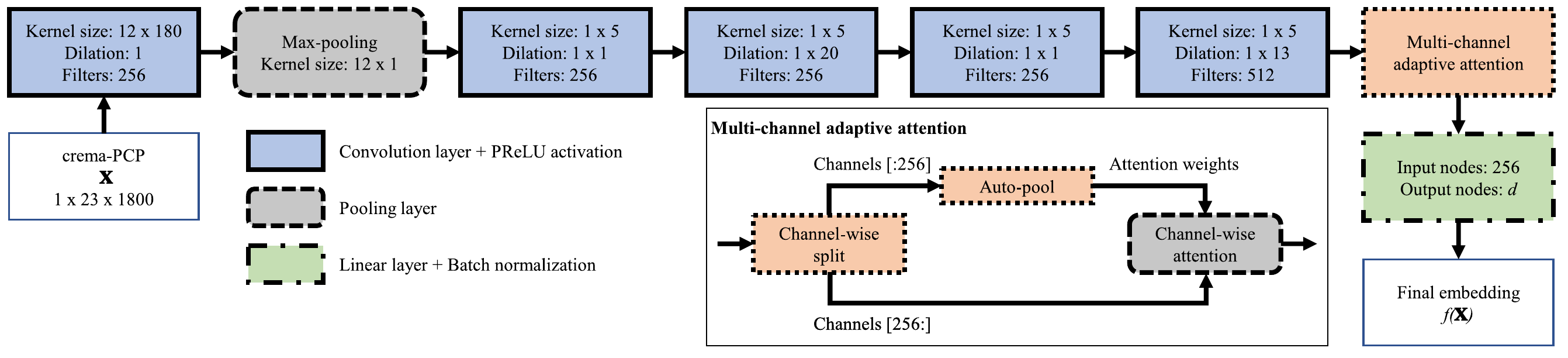}
%  \vspace{2.0cm}
  \caption{Block diagram of \modelname's architecture.}
  \label{fig:model}
\end{figure*}

\section{Musically-motivated version embeddings}
\label{sec:model}

\subsection{Input}

We use as input a relatively novel PCP variant: crema-PCP. This representation is constructed by using the output of an intermediate step of the crema chord estimation model~\cite{mcfee2017structured}. For each frame, the crema model estimates the root, the bass, and the pitch classes, which are later combined to output a single chord. Specifically, crema-PCP is constructed by taking the sigmoid activation values of pitch classes for each frame, and considering them as the energy values of each pitch class~\cite{mcfee2017structured}. Although being a fairly new approach, crema-PCP has been shown to outperform elaborate PCP representations in some benchmarking experiments~\cite{yesiler2019}. We use the pre-trained model available at \url{https://github.com/bmcfee/crema} (version 0.1.0) and denote the obtained output by $\bm{X}\in [0,1]^{12\times T}$, where $T$ is the number of frames using non-overlapping windows of 93\,ms. For training, we take random patches of $T=1800$ frames after applying data augmentation (see below) to a full song. At inference time, we give entire tracks to the model without picking random patches of a particular length (preliminary experiments showed that the below-proposed temporal pooling strategy was also effective with entire tracks at inference time).

\subsection{Network architecture}
\label{sec:architect}

\modelname\ consists of 5~convolutional blocks with PReLU activation functions and no padding, interleaved by two different pooling layers (Figure~\ref{fig:model}). A linear layer followed by non-parametric batch normalization produces the final embedding. With the current best setup, the total number of parameters is 6.3\,M. We now motivate and present the key components of \modelname.

\noindent
\textbf{Transposition-invariant architecture ---} Following the strategy proposed by Xu~et~al.~\cite{xu2018}, we increase the dimension of the crema-PCP inputs $\bm{X}$ from 12$\times T$ to 23$\times T$ by concatenating two copies of $\bm{X}$ in the pitch dimension and removing the last pitch class. The first convolutional layer, with a kernel size of 12$\times$180 traverses the input, going through all possible transpositions in the pitch dimension, and the subsequent max-pooling layer, with a kernel size of 12$\times$1, keeps the transposition with the highest activation value (convolutions in \modelname\ have no padding). 

\noindent
\textbf{Expanding the receptive field ---} The 4~convolutional blocks after max-pooling are designed to encode higher-level information and to increase the receptive field of the model (Figure~\ref{fig:model}). On the one hand, with the layers that have no dilation, we aim to encode higher-level nonlinearities without expanding the temporal context. On the other hand, with the layers that have dilations 20 and 13, we increase the receptive field, which after max-pooling is less than 17\,s, to approximately 30\,s. Notice that this temporal span could be already sufficient to detect musical piece versions, at least from a human perspective. However, to process an even larger time span, and to be able to deal with different lengths $T$ at test time, we still perform an additional step.

\noindent
\textbf{Summarizing temporal content ---} We consider the convolutional part of our network as a feature extractor that processes the input to obtain a representation that is invariant to the modifiable musical characteristics mentioned in Section~\ref{sec:intro}. In order to summarize the values of each feature in the temporal dimension, unlike previous approaches that use average- or max-pooling variants~\cite{doras2019, yu2019}, we propose a multi-channel adaptive attention mechanism, which combines multi-channel temporal attention~\cite{Serra18AIRD} with auto-pool~\cite{mcfee2018}. The first idea is to let the network compute (and learn) the importance of each time step independently for each feature with an attention-like mechanism~\cite{Serra18AIRD}. The second idea is to apply a non-linear, learnable pooling function with a scaling parameter before the softmax function~\cite{mcfee2018} such that, depending on the value of such parameter, the function pivots between average- and max-pooling. In practice, temporal summarization is done by calculating channel-wise attention weights, which correspond to the first half of the filters of the last convolutional layer, using the auto-pool function, and utilizing the result to weight the last half of the filters of the same layer. Splitting the hidden representation channel-wise into two halves, $\bm{H}=[\bm{H}_a~\bm{H}_b]$, this corresponds to
\begin{equation*}
\bm{H}' = \sum_{t=1}^{T} \sigma(\alpha\bm{H}_a) \odot \bm{H}_b ,
\end{equation*}
where the sum is taken across the temporal dimension, $\sigma$ corresponds to the softmax function, $\alpha$ is a learnable parameter which we initialize to 0 (equivalent to average-pooling), and $\odot$ is the element-wise product.

\noindent
\textbf{Standardizing embedding components ---} For deep metric learning approaches using a triplet loss, it is highly important to take into account the volume of the hyper-dimensional space where the embeddings lie, specially during training. For instance, if the magnitude of the distances and the margin are disproportionate, the training process may not be able to structure the latent space in an effective way. With these motivations in mind, we propose to use non-parametric batch normalization after the linear layer that finalizes the encoding process. By doing so, we aim to obtain zero-mean and unit-variance components in our embeddings, yielding a statistically-standardized latent space volume. This, together with dimension-normalized Euclidean distances, may also allow us to develop some intuition regarding the loss values and the corresponding margin.

\subsection{Training strategy}\label{subsec:training_strategy}

\modelname\ is trained by minimizing the triplet loss
\begin{equation}\label{eq:triplet}
L = \max \left( D\left(\bm{X}_A,\bm{X}_P\right) - D\left(\bm{X}_A,\bm{X}_N\right) + m, 0\right) 
\end{equation}
using
\begin{equation*}\label{eq:dists}
D(\bm{X}_i,\bm{X}_j) = \frac{1}{d}~ \lVert f(\bm{X}_i) - f(\bm{X}_j)\rVert^2 ,
\end{equation*}
where $\lVert~\rVert$ corresponds to the Euclidean norm and  $f(\bm{X})$ denotes an embedding of size $d$ produced by our model. Equation~\ref{eq:triplet} aims to make the distance between an anchor $A$ and a positive example $P$ smaller than the distance between the same anchor $A$ and a negative example $N$ under a margin $m$. We now present our decisions regarding training data, data augmentation, triplet mining, and hyper-parameters.

\noindent
\textbf{Training data ---} We use a private collection of 97,905 songs that are divided into 17,999 cliques. The annotations of the songs are under the Creative Commons BY-NC 3.0 license, and obtained with the API of \url{secondhandsongs.com}. The related metadata can be found at our repository. For training and validation, we created two disjoint sets of cliques, with 14,499 cliques containing 83,905 songs and 3,500 cliques containing 14,000 songs, respectively. All audio files are encoded in MP3 format and their sample rate is 44.1\,kHz.

\noindent
\textbf{Data augmentation ---} In order to enhance the learning of \modelname, we apply to each example a data augmentation function specifically designed for VI. Based on the modifiable musical characteristics specified in Section~\ref{sec:intro} and elsewhere, such function sequentially and independently applies transposition in the pitch dimension, time stretching, and time warping with probabilities 1, 0.3, and 0.3, respectively. Transposition uses the octave-equivalent characteristics of PCP representations and randomly rolls $\bm{X}$ in the pitch dimension between 0 and 11~bins. Time stretching uses one-dimensional interpolations in the temporal domain, with a random factor between 0.7 and 1.5. Time warping consists of three mutually-exclusive functions, which either silence, duplicate, or remove frames with probabilities 0.3, 0.4, and 0.3, respectively (silence corresponds to zeroing-out the entire frame). Once selected, these functions are applied on a per-frame basis with a probability of 0.1, 0.15, and 0.1, respectively. All random numbers are sampled using a uniform distribution. %In our analysis, we prove that this data augmentation method improves the generalization performance of the network significantly.

\noindent
\textbf{Triplet mining ---} As discussed in previous works that employ a triplet loss, the characteristics of the triplets in each mini-batch may have drastic effects on learning performance~\cite{schroff2015, hermans2017}. For our model, we employ an online hard triplet mining strategy~\cite{hermans2017}. In our implementation, we choose 16 unique cliques and 4 songs per clique, forming a mini-batch of 64. For the cliques that have less than 4 songs, we choose among the already chosen songs of the same clique. Within a mini-batch, we consider all the examples as anchors ($A$), and select the positive/negative example that has the maximum/minimum distance to the anchor ($P$ and $N$, respectively; Equation~\ref{eq:triplet}). Although Schroff~et~al.~\cite{schroff2015} point out that the hardest examples may lead to local minima early in the training, our triplets can be considered ``moderate''~\cite{hermans2017}, as they are selected only from the current mini-batch, and therefore do not strictly correspond to the hardest triplets in the dataset. This presumably avoids the aforementioned local minima.

\noindent
\textbf{Hyper-parameters and optimization ---} We train our network for 120~epochs with plain stochastic gradient descent, using an initial learning rate of 0.1 and decreasing it by a factor of 5 at epochs 80 and 100. An epoch is completed when our data loader goes through all possible cliques. However, an important detail to note is that we include the cliques with size between 6 and 9 twice, the ones with size between 10 and 13 three times, and the ones with size 14 or above four times. This is done to increase the probability of every song being introduced to the network at least once per epoch. The margin value $m$ for the triplet loss is 1. As mentioned, we use patches of $T=1800$ frames for training and an initial auto-pool parameter $\alpha=0$. If not already specified in Figure~\ref{fig:model}, the remaining hyper-parameters and implementation details can be found at our GitHub repository~\footnote{\url{https://github.com/furkanyesiler/move}}. We study the impact of the embedding dimension $d$ in the next section.
\vspace{-2mm}
\section{Results}
\label{sec:results}

\subsection{Evaluation methodology}

For studying the effect of the embedding dimension and performing the ablation study, we train with a subset of our training set with 8,817~cliques and 44,909~songs in total, and report the performance scores on our validation set. For comparison to the previous work, we utilize the entire training set. To report performance, we use mean average precision (MAP) and mean rank of the first relevant result (MR1). For all the experiments presented in this section, we use the models obtained after the last epochs.

For comparing the performance of \modelname\ with the state of the art, we use two additional datasets. The first dataset, the benchmark subset of Da-TACOS~\cite{yesiler2019}, contains a total of 15,000~songs, with 1,000~cliques of 13~songs each, and 2,000~songs not belonging to any other clique (acting as noise and not queried). The second dataset, YouTubeCovers (YTC)~\cite{silva2015}, contains 50~cliques with 7~songs each, and comes split into a training and a test set with 250 and 100~songs each, respectively. To compare the performance of our model on YTC with previous works, we follow their approach of only querying the test set to retrieve the versions in the reference set~\cite{silva2016, seetharaman2017, xu2018, yu2019}. Moreover, in this case, we remove from our training data the 17~cliques that overlap with YTC. After that, both Da-TACOS and YTC do not contain any overlapping cliques with respect to our training/validation data.
\vspace{-1mm}

\subsection{Effect of the embedding dimension}

\begin{figure}[tb!]
  \centering
  \includegraphics[width=\columnwidth]{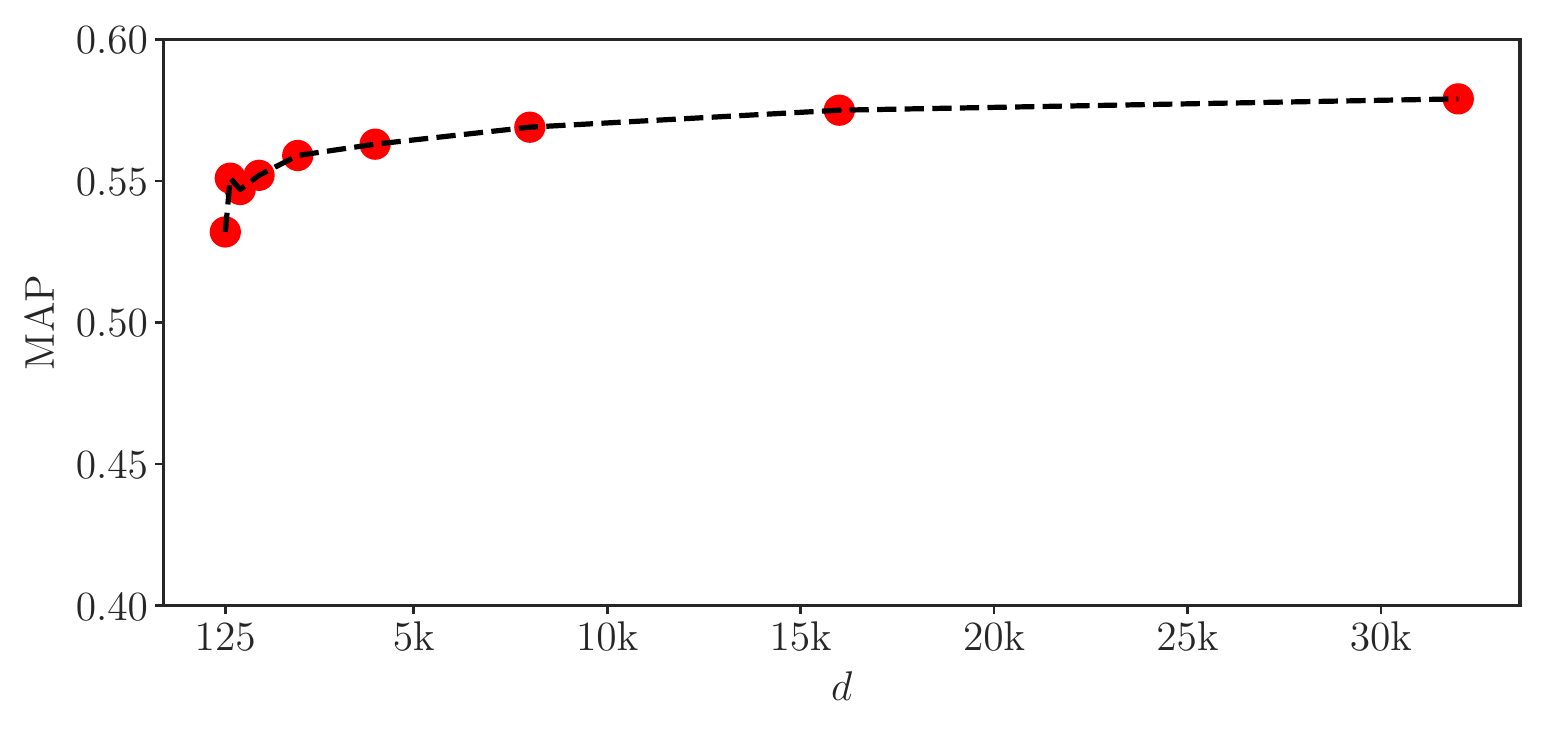}
  \vspace{-7mm}
  \caption{MAP with respect to embedding dimension $d$ on validation data.}
  \label{fig:embmap}
  \vspace{-2mm}
\end{figure}

For any embedding system, the size of the embeddings $d$ is a crucial hyper-parameter, as it can have an important effect on model performance. Therefore, we decide to study the model performance on the validation set with respect to it (Figure~\ref{fig:embmap}). For this set of experiments, we consider $d=\{128, 256, 512, 1\,\text{k}, 2\,\text{k}, 4\,\text{k}, 8\,\text{k}, 16\,\text{k},\linebreak 32\,\text{k}\}$. We observe that performance continues to increase with the embedding dimensionality until it saturates at $d=16\,\text{k}$. We can place a knee in the curve between $d=512$ and $d=2\,\text{k}$.

\vspace{-2mm}

\subsection{Ablation study}

\begin{table}[tb!]
\setlength\tabcolsep{10pt}
\begin{center}
\label{tab:ablation}
\begin{tabular}{l c c}
\hline\hline
 & MAP &  MR1 \\
\hline\hline

\modelname & 0.575  &  156  \\ % 0.574 & 156

\hline%\hline
\multicolumn{3}{c}{\textit{Data augmentation}} \\ 
1: Without data augmentation & 0.540   &  180  \\ % 0.538 & 195
\hline%\hline
%\multicolumn{3}{c}{\textit{Input size}} \\  
%2: Random length  &   &  \\
%\hline%\hline
\multicolumn{3}{c}{\textit{Transposition invariance}} \\ 
2: Without transposition invariance &    0.154 & 399 \\ % 0.069 & 1896
\hline%\hline
\multicolumn{3}{c}{\textit{Summarizing temporal content}} \\ 
3: Only multi-channel attention &  0.575 &  153  \\ % 0.575 & 155
4: Only auto-pool &  0.563  & 145   \\ % 0.560 % 150
5: Max-pooling & 0.561  &  152 \\ % 0.557 % 153
6: Average-pooling & 0.491  &  197 \\ % 0.495 & 203
\hline%\hline
\multicolumn{3}{c}{\textit{Triplet mining strategies}} \\
7: Semi-hard mining & 0.545  & 135 \\ % 0.548 & 140
8: Random mining & 0.427  & 167 \\ % 0.429 & 166

\hline\hline
\end{tabular}
\caption{Ablation study. Performance on the validation set using $d=16\,\text{k}$.}\label{tab:ablation}
\end{center}
\tabspace
\end{table}

We now analyze the performance of the main components of our network by comparing them to their potential alternatives (Table~\ref{tab:ablation}). With that, we aim to quantify the importance of each decision. The first aspect we assess is the effect of the proposed data augmentation strategy (1). We find that removing data augmentation yields a relative decrease of 6\% in MAP. The second aspect that we evaluate is the importance of the transposition-invariant architecture explained in Section~\ref{sec:architect} (2). As an alternative, we consider the case where we do not pre-process the input by changing its shape, and remove the max-pooling layer after the first convolution. Although trained with a much smaller learning rate ($10^{-4}$) and the Adam optimizer, the model was not able to properly learn an effective representation, even though multiple transpositions were present in the data augmentation function. The third aspect we consider is temporal summarization (3--6). We observe that the introduction of the auto-pool parameter $\alpha$ to multi-channel attention does not really change the results (3). In contrast, substituting the proposed multi-channel attention by auto-, max-, or average-pool clearly has an impact (4--6). The final aspect we analyze is the effect of the triplet mining strategy (7--8). To do so, we train our network with online semi-hard (7) and random (8) mining strategies. For semi-hard mining, we pick a random positive example for each anchor and then select a negative example that satisfies the condition $D\left(\bm{X}_A,\bm{X}_N\right) \leq D\left(\bm{X}_A,\bm{X}_P\right)$. In case no such negative example exists, we pick a random one. For random mining, we randomly select one positive and one negative example for each anchor. We see that semi-hard and random mining produce a relative MAP decrease of 5 and 26\%, respectively. Overall, our ablation study shows that all introduced variations have a positive impact in performance. The only exception is the mixing of the auto-pool parameter with multi-channel attention, which nonetheless does not substantially affect the performance.

\subsection{Comparison with the state-of-the-art}

Finally, we compare the performance of \modelname\ with the state of the art (Table~\ref{tab:sota}). The results on Da-TACOS show that \modelname\ clearly outperforms all considered VI systems. Importantly, this does not only happen for systems that, like \modelname, use a single input representation and alignment, but also for complex systems that employ early or late fusion strategies. The relative MAP difference with respect to LateFusion~\cite{chen2018}, the most competing system, is over 10\%. We also see that, although the best performance is achieved with a relatively large embedding dimension of 16\,k, a smaller embedding size of 4\,k can still outperform the state of the art. %We would like to note that the results for the previous approaches were obtained using a lower temporal resolution (non-overlapping windows of 465\,ms) for the feature representations.
The results on YTC support the claim that \modelname\ achieves a new state-of-the-art performance (Table~\ref{tab:sota}). However, we caution about the use of YTC to report VI performance, as differences measured with this dataset may not be significant due to the relatively small number of query and reference tracks (cf.~\cite{serra2011}). As an example, \modelname\ with $d=4\,\text{k}$ shows a similar result as the setting with $d=16\,\text{k}$ on YTC, while in larger datasets, the latter clearly outperforms the former.

\begin{table}[tb!]
\setlength\tabcolsep{12pt}
\begin{center}
\label{tab:sota}
\begin{tabular}{l c c}
\hline\hline
 & MAP &  MR1\\
\hline\hline

\multicolumn{3}{c}{\textit{Results on Da-TACOS}}\\

%\hline%\hline
2DFTM~\cite{bertin2012}  & 0.275  & 155 \\

SiMPle~\cite{silva2016}  &  0.332 &  142\\

Dmax~\cite{chen2018}  & 0.322 & 132 \\

Qmax~\cite{serra2009}  & 0.365   & 113  \\

Qmax*~\cite{serra2009unsupervised}  &  0.373 & 104 \\

EarlyFusion~\cite{tralie2017cover}  & 0.426  &   116 \\

LateFusion~\cite{chen2018}  & 0.454  &   177 \\

\modelname\ w/ $d=4\,\text{k}$ (ours) &  0.495 & 42 \\ % 0.489  & 43 \\ 

\modelname\ w/ $d=16\,\text{k}$ (ours) & \textbf{0.507} & \textbf{40} \\ % \textbf{0.506}  & \textbf{42} \\ 
\hline%\hline
\multicolumn{3}{c}{\textit{Results on YTC}} \\ 
%\hline%\hline

SiMPle~\cite{silva2016}  &  0.591   & 8 \\

2DFTM sequences~\cite{seetharaman2017}  & 0.648    &  8 \\

InNet~\cite{xu2018}  & 0.660  &  6 \\

SuCo-DTW~\cite{silva2018}  & 0.800  & \textbf{3} \\

CQT-TPPNet~\cite{yu2019}  & 0.859    & \textbf{3} \\

\modelname\ w/ $d=4\,\text{k}$ (ours) &  \textbf{0.889} & \textbf{3} \\ 

\modelname\ w/ $d=16\,\text{k}$ (ours) & 0.888 & \textbf{3} \\ % \textbf{0.885}  & \textbf{3}  \\ 
\hline\hline
\end{tabular}
\caption{Comparison of state-of-the-art VI systems (best results are highlighted in bold). Results on Da-TACOS are taken from~\cite{yesiler2019}.}\label{tab:sota}
\end{center}
\tabspace
\vspace{-1mm}
\end{table}
\section{Conclusion}

In this work, we have proposed \modelname, a method for accurate and scalable version identification using musically-motivated embeddings. \modelname\ achieves state-of-the-art performance on two publicly-available benchmark sets for VI. After motivating the components of its architecture and training strategy, both designed while incorporating a certain degree of domain knowledge, we performed an ablation study to justify our decisions. We have also studied the relation between the embedding size and the performance of our model. As future work, we plan to investigate different input representations. Since some early and late fusion methods incorporate several musical dimensions to outperform their isolated components, we intend to explore possibilities where we can mimic the same idea to improve \modelname's performance. Moreover, considering that our method outperforms traditional VI systems that are built with a certain notion of similarity in mind (for instance, local alignment between tonal features), a future study investigating the similarity concept learned by our model could provide meaningful insight regarding the links that bind various versions originated from the same musical composition.

\section{Acknowledgments}
This work is supported by the MIP-Frontiers project, the European Union's Horizon 2020 research and innovation programme under the Marie Skłodowska-Curie grant agreement No.~765068, and by TROMPA, the Horizon 2020 project 770376-2.

% References should be produced using the bibtex program from suitable
% BiBTeX files (here: strings, refs, manuals). The IEEEbib.bst bibliography
% style file from IEEE produces unsorted bibliography list.
% -------------------------------------------------------------------------
\clearpage
\bibliographystyle{IEEEbib}
\bibliography{strings,refs}

\end{document}